\documentclass[12pt,showpacs,preprintnumbers,amsmath,amssymb]{revtex4}

\usepackage{graphicx}
\usepackage{dcolumn}
\usepackage{bm}

\begin{document}


\title{Variants of Bell inequalities}
\author{Zeqian Chen}
\email{chenzq@mail.cnu.edu.cn}
\affiliation{%
Research Center for Mathematical Physics, School of Mathematical
Sciences, Capital Normal University, 105 North Road, Xi-San-Huan,
Beijing, China}
\affiliation{%
State Key Laboratory of Magnetic Resonance and Atomic and Molecular
Physics, Wuhan Institute of Physics and Mathematics, Chinese Academy
of Sciences, 30 West District, Xiao-Hong-Shan, P.O.Box 71010, Wuhan
430071, China}

\date{\today}

\begin{abstract}
A family of Bell-type inequalities is present, which are constructed
directly from the ``standard" Bell inequalities involving two
dichotomic observables per site. It is shown that the inequalities
are violated by all the generalized Greenberger-Horne-Zeilinger
states of multiqubits. Remarkably, our new inequalities can provide
stronger non-locality tests in a sense that the local reality
inequalities are exponentially stronger than the corresponding
multipartite separability inequalities. This reveals that the
exponential violation of local realism by separable states is an
interesting consequence of quantum fluctuation of multipartite
systems.
\end{abstract}

\pacs{03.67.Mn, 03.65.Ud}
\maketitle

\newpage

Bell's inequality \cite{Bell} was originally designed to rule out
various kinds of local hidden variable theories based on Einstein,
Podolsky, and Rosen's (EPR's) notion of local realism \cite{EPR},
and thus certifies quantum origin and true nonlocality of entangled
states \cite{Mermin}. Bell inequalities now serve a dual purpose. On
one hand these inequalities provide a test to distinguish entangled
from non-entangled quantum states. Indeed, experimenters routinely
use violations of Bell inequalities to check whether they have
succeeded in producing entangled states \cite{GN}. On the other
hand, violations of the inequalities are applied to realizing
certain tasks by quantum information, such as building quantum
protocols to decrease communication complexity \cite{BZZ} and making
secure quantum communication \cite{SGA}. Derivations of new and
stronger Bell-type inequalities are thus one of the most important
and challenging subject in quantum theory.

There are extensive earlier works on Bell inequalities \cite{arxiv},
including Clauser-Horne-Shimony-Holt (CHSH) inequality for bipartite
systems \cite{CHSH} and Mermin-Roy-Singh-Ardehali-Belinskii-Klyshko
(MK) inequalities for multi-particle systems \cite{MK}. A set of
multipartite Bell inequalities with two dichotomic observables per
site, including the MK inequalities, has been constructed by Werner
and Wolf and by \.{Z}ukowski and Brukner (WWZB) \cite{WWZB}, which
is complete in the sense that the inequalities are satisfied if and
only if the correlations considered are describable in a local and
realistic picture. Such inequalities are referred usually as
``standard" ones. There are also other Bell-type inequalities with
more than two observables per site and/or for high dimensional
systems \cite{NSBell}, quadratic Bell inequalities \cite{Uffink},
and ones based on variances and Wigner-Yanase skew information
\cite{Chen}. We refer to \cite{review} and references therein for
more details.

The theorem of Gisin \cite{G} states that any pure entangled state
of two qubits violates a CHSH inequality. Can Gisin's theorem be
generalized to multiqubits? Scarani and Gisin \cite{SG} show that
there are states which do not violate the MK inequalities.
Surprisingly, \.{Z}ukowski {\it et al} \cite{ZBLW} further show that
there is a family of pure entangled states of $n>2$ qubits which do
not violate any standard Bell inequality. This family is a subset of
the generalized GHZ states given by\begin{equation}| \Psi (\theta)
\rangle = \cos \theta |0^n \rangle + \sin \theta | 1^n \rangle,
\end{equation}with $0 \leq \theta \leq \pi /4.$ The GHZ state \cite{GHZ}
is for $\theta = \pi /4.$ However, K.Chen, S.Albeverio and S.-M.Fei
\cite{CAF} presented recently a family of Bell inequalities
involving only two measurement settings per observer and show by an
analytical proof that all the generalized GHZ entangled states of
$n>2$ qubits can violate the inequalities \cite{CWKO}. Such
inequalities are highly desirable because they can lead to a much
easier and more efficient way to test nonlocality, and contribute to
the development of novel quantum protocols and cryptographic schemes
by exploiting much less entangled resources and experimental
efforts.

Here, we present a family of Bell-type inequalities, which are
variants of the ``standard" Bell inequalities. It is proved that all
the generalized GHZ entangled states given by Eq.(1) can violate the
inequalities. For $n$ qubits, the quantum separability bound of our
new inequalities is $2^{n-1},$ while the maximum possible
entanglement violation is $2^{(n-1)/2}+ 2^{n-1},$ which can be
archived only by the GHZ state. Remarkably, the bound of local
realism of the inequalities is $2,$ independent of $n$ qubits. This
yields that the local reality inequalities are exponentially
stronger than the corresponding multipartite separability
inequalities. Indeed, for $n>2$ qubits the inequalities show that
separable quantum states do not satisfy the local reality
inequalities, in contrast to the widespread opinion that any
separable quantum state satisfies every classical probabilistic
constraints. This is so because the inequalities are based on both
mean values and variances of observables and hence exhibit effects
of quantum fluctuation of multipartite systems. Thus, our new
inequalities demonstrate nonclassical properties of separable
quantum states and reveal that, the exponential violation of local
realism by separable quantum states is an interesting consequence of
quantum fluctuation of multipartite systems.

Let us give a brief review of MK Bell operators and the associated
MK inequalities. The MK Bell operators of $n$ qubits are defined
recursively ($n \geq 2$). Let
$\vec{a}_j\vec{\sigma}_j,\vec{a}'_j\vec{\sigma}_j$ denote spin
observables on the $j$-th qubit, $j=1,...,n,$ where all $\vec{a}_j,
\vec{a}'_j$ are unit vectors in $\mathbb{R}^3$ and $\vec{\sigma}_j =
(\sigma^j_x, \sigma^j_y, \sigma^j_z)$ is the Pauli matrices on the
$j$-th qubit. Denote by ${\cal B}_1 = \vec{a}_1 \vec{\sigma}_1$ and
${\cal B}'_1 = \vec{a}'_1 \vec{\sigma}_1.$
Define\begin{equation}{\cal B}_n = {\cal B}_{n-1} \otimes
\frac{1}{2}(\vec{a}_n\vec{\sigma}_n + \vec{a}'_n\vec{\sigma}_n) +
{\cal B}'_{n-1} \otimes \frac{1}{2}( \vec{a}_n\vec{\sigma}_n -
\vec{a}'_n\vec{\sigma}_n ),\end{equation}where ${\cal B}'_n$ denotes
the same expression ${\cal B}_n$ but with all the $\vec{a}_j$ and
$\vec{a}'_j$ exchanged. ${\cal B}_n$ is called the MK Bell operator
of $n$ qubits. Assuming ``local realism" \cite{EPR,Bell}, one
concludes the MK inequality of $n$ qubits as
follows:\begin{equation}\langle {\cal B}_n \rangle \leq
1,\end{equation}which can be violated by quantum mechanics
\cite{MK}.

By convention, we adopt the notation $|0^n \rangle = |0 \cdots 0
\rangle$ and $|1^n \rangle = |1 \cdots 1 \rangle.$ Recall that when
all the $\vec{a}_j$ and $\vec{a}'_j$ are in the $x-y$ plane and
$\vec{a}_j$'s are distributed with angles $(j-1)(-1)^{n+1}2 \pi /
(2n)$ with respect to the $x$-axis and $\vec{a}_j \perp \vec{a}'_j,$
the associated MK Bell operator has the following spectral
decomposition \cite{SG}:\begin{equation}{\cal M}_n =
2^{(n-1)/2}\left ( |\textrm{GHZ}_+\rangle \langle \textrm{GHZ}_+ | -
|\textrm{GHZ}_-\rangle \langle \textrm{GHZ}_- | \right
),\end{equation}where $|\textrm{GHZ}_{\pm} \rangle =
\frac{1}{\sqrt{2}}\left ( |0^n \rangle \pm |1^n \rangle \right )$
are GHZ's states \cite{GHZ}. Set\begin{equation}{\cal V}_n = {\cal
M}_n + {\cal M}^2_n.\end{equation}Then,\begin{equation}\langle {\cal
V}_n \rangle = \langle {\cal M}_n \rangle + \langle {\cal M}_n
\rangle^2 + \Delta ( {\cal M}_n ),\end{equation}where $\Delta (
{\cal M}_n ) = \langle ({\cal M}_n - \langle {\cal M}_n \rangle )^2
\rangle$ is the variance of ${\cal M}_n$ in a state.

{\it The separability inequalities for $n$ qubits}.---For any
separable quantum state of $n$ qubits, one
has\begin{equation}\langle {\cal V}_n \rangle \leq
2^{n-1}.\end{equation}Indeed, note that ${\cal V}_n = 2^{(n-1)/2}
(|0^n \rangle \langle 1^n | + |1^n \rangle \langle 0^n |) +
2^{n-1}(|0^n \rangle \langle 0^n | + |1^n \rangle \langle 1^n |).$
Then, for every product state $| \psi \rangle = | \psi_1
\rangle\cdot\cdot\cdot| \psi_n \rangle$ of $n$ qubits, one
has$$\begin{array}{lcl}\langle {\cal V}_n \rangle &=& 2^{(n-1)/2}
(\prod^n_{j=1} \alpha_j \beta^*_j + \prod^n_{j=1} \alpha^*_j
\beta_j)\\&~& + 2^{n-1} (\prod^n_{j=1} |\alpha_j|^2 +
\prod^n_{j=1} |\beta_j|^2)\\
& \leq & 2^{(n-1)/2} (\prod^n_{j=1} |\alpha_j| + \prod^n_{j=1}
|\beta_j|)^2\\&~& + (2^{n-1}- 2^{(n-1)/2} )(\prod^n_{j=1}
|\alpha_j|^2 + \prod^n_{j=1} |\beta_j|^2),\end{array}$$where
$\alpha_j = \langle \psi_j | 0 \rangle$ and $\beta_j = \langle
\psi_j | 1 \rangle,$ $j=1,\ldots, n.$ Using $\max (x \sin \phi + y
\cos \phi ) = \sqrt{ x^2 + y^2},$ we get $(\prod^n_{j=1} |\alpha_j|
+ \prod^n_{j=1} |\beta_j|)^2 \leq 1$ for $n \geq 2.$ This concludes
that Eq.(7) holds for all product states. Since a separable state is
a convex combination of product states, it is concluded that Eq.(7)
holds for all separable quantum states of $n \geq 2$ qubits.

On the other hand, for every product state $| \psi \rangle = |
\psi_1 \rangle\cdot\cdot\cdot| \psi_n \rangle,$ there is a local
unitary transformation $U=U_1 \otimes \cdot\cdot\cdot \otimes U_n$
such that $U | \psi \rangle = |0^n \rangle$ and so, $\langle \psi |
U^{\dagger} {\cal V}_n U | \psi \rangle = 2^{n-1},$ where
$U^{\dagger}$ denotes the adjoint operator of $U.$ This yields that
the equality of Eq.(7) can be archived by product states.

{\it The maximum possible entanglement violation.}---By Eq.(4), for
any entangled state we have\begin{equation}\langle {\cal V}_n
\rangle \leq 2^{(n-1)/2} + 2^{n-1}.\end{equation}It is easy to check
that for the generalized GHZ states $| \Psi (\theta) \rangle$ of
Eq.(1),\begin{equation} \langle \Psi (\theta) | {\cal V}_n | \Psi
(\theta) \rangle = 2^{(n-1)/2} \sin 2 \theta +
2^{n-1}.\end{equation}Thus, all generalized GHZ entangled states
violate Eq.(7) and, the GHZ state is the only state that violates
Eq.(7) maximally because the GHZ state has been shown to be the only
state that violates the MK inequality maximally \cite{SG,Chen2}.

{\it The local reality inequalities}.---From the classical view of
local realism, the values of
$\vec{a}_j\vec{\sigma}_j,\vec{a}'_j\vec{\sigma}_j$ are predetermined
by a local hidden variable (LHV) $\lambda$ before measurement, and
independent of measurements, orientations or actions performed on
other parties at spacelike separation. We denote by $\varrho
(\lambda)$ the statistical distribution of $\lambda$ satisfying
$\varrho (\lambda) \geq 0$ and $\int d \lambda \varrho (\lambda)
=1.$ Since $-1 \leq {\cal M}_n ( \lambda ) \leq 1$ for the local
hidden variable $\lambda,$ one has$$\langle {\cal M}_n
\rangle_{\mathrm{LHV}} = \int d \lambda \varrho (\lambda){\cal M}_n
( \lambda ) \leq 1,$$and$$\begin{array}{lcl}\Delta ( {\cal M}_n
)_{\mathrm{LHV}} &=& \int d \lambda \varrho (\lambda)\left [ {\cal
M}_n ( \lambda ) - \langle {\cal M}_n \rangle_{\mathrm{LHV}} \right
]^2\\& \leq & 1 - \langle {\cal M}_n
\rangle^2_{\mathrm{LHV}}.\end{array}$$Therefore, by Eq.(6) we
have\begin{equation}\langle {\cal V}_n \rangle_{\mathrm{LHV}} =
\langle {\cal M}_n \rangle_{\mathrm{LHV}} + \langle {\cal M}_n
\rangle^2_{\mathrm{LHV}} + \Delta ( {\cal M}_n )_{\mathrm{LHV}} \leq
2.\end{equation}Surprisingly, combining the separability inequality
Eq.(7) and the local reality inequality Eq.(10) we conclude that
separable quantum states of $n > 2$ qubits do not satisfy the local
reality inequalities, in contrast to the widespread opinion that any
separable quantum state satisfies every classical probabilistic
constraints. That separable quantum states do not satisfy the local
reality inequalities has been pointed out by Loubenets
\cite{Loubenets}, but her definition of classicality involved is
narrower than the usual concept of LHV and consequently,
Ref.\cite{Loubenets} does not demonstrate nonclassical properties of
separable states \cite{Simon}. However, our analysis based on both
the mean values and variances of observables does involve the usual
sense of LHV and indicates that the violation of local realism by
separable states is an interesting consequence of quantum
fluctuation of multipartite systems.

As already demonstrated by Werner \cite{Werner}, testing for
entanglement within quantum theory, and testing quantum mechanics
against LHV theories are not equivalent. Indeed, as shown in
Ref.\cite{Roy}, the physical origins of EPR's local realism and
quantum entanglement are different. For a multipartite system which,
having interacted in past, are now spatially separated, EPR's local
realism means that elements of physical reality for one subsystem
should be independent of what is done with the others. In contrast,
quantum entanglement refers only to quantum multipartite states,
whether or not the individual subsystems are spatially separated.
The violation of Bell's inequalities assuming EPR's local realism by
suitable entangled states is therefore an interesting but indirect
consequence of quantum entanglement. Further, our Bell-type
inequalities show that quantum fluctuation of multipartite systems,
even in product states, can exhibit quantum nonlocality against LHV
theories based on EPR's local realism.

In summary, we have presented a family of Bell-type inequalities,
which are constructed directly from the ``standard" Bell
inequalities involving two dichotomic observables per site. It is
shown that the inequalities are violated by all the generalized GHZ
entangled states of multiqubits. Remarkably, our new inequalities,
based on both mean values and variances of observables, can provide
stronger non-locality tests in a sense that the local reality
inequalities are exponentially stronger than the corresponding
multipartite separability inequalities. This reveals that the
exponential violation of local realism by separable states is an
interesting consequence of quantum fluctuation of multipartite
systems. Complementary to the standard inequalities and a number of
existing results, our result furthermore shed considerable light on
LHV theories based on EPR's local realism. Evidently, the argument
involved here can be generalized to all standard Bell inequalities.
We hope that those variants of the usual Bell inequalities will play
an important role in quantum information.

This work was supported by the National Natural Science Foundation
of China under Grant No.10571176, the National Basic Research
Programme of China under Grant No.2001CB309309, and also funds
from Chinese Academy of Sciences.


\end{document}